\documentclass[10pt]{article}
\newlength{\vshift}
\input epsf.tex
\newlength{\hshift}

\usepackage{amssymb}
\usepackage{amsmath}
\usepackage{amsthm}
\usepackage{amsopn}
\allowdisplaybreaks
\usepackage{color}

\begin{document}

{\textsf{\today}
 \vspace*{3cm}

\begin{center}
{\large \bf{ Generatoin of Circular Polarization of CMB via Polarized Compton Scattering}}
\vskip 2em
{
 {\bf Ali Vahedi} \footnote{vahedi@khu.ac.ir } ,
{\bf Jafar Khodagholizadeh }\footnote{gholizadeh@ipm.ir},
{\bf Rohoollah Mohammadi} \footnote{ rmohammadi@ipm.ir}
{and \bf Mahdi Sadegh} \footnote{ m.sadegh@pnu.ac.ir,}
}
\vskip 1em {\it  1 Department of Physics, Kharazmi University, Mofateh Ave, P.O. Box 15614, Tehran, Iran.}
\vskip 1em {\it 2 Farhangian University, P.O. Box 11876-13311, Tehran, Iran.}
\vskip 1em {\it  3 Iranian National Museum Of Science and Technology  (INMOST), PO BOX: 11369-14611, Tehran, Iran.\\ School of Astronomy, Institute for Research in Fundamental Sciences (IPM), P. O. Box 19395-5531, Tehran, Iran.}
\vskip 1em {\it  4 Department of Physics, Payame Noor University (PNU), P.O. Box 19359-3697, Tehran, Iran.}
\end{center}
 \vspace*{0.5cm}
\begin{abstract}
The standard scenario of cosmology predicts a measurable amount for
linear polarization of the Cosmic Microwave Background radiation (CMB) via Thomson scattering, while through this scenario, the generation of circular polarization is excluded. On the
another hand, the circular polarization of CMB has not been excluded in observational evidence. The generation of CMB photons circular polarization via their Compton scattering with polarized cosmic electrons is considered in this paper. Our motivation for considering polarized Compton scattering comes from the effects of the external magnetic field in large scale, the chiral magnetic instability and new physics interactions of the cosmic electrons. It is shown that damping term of polarized Compton scattering in the presence of scalar perturbation can generate circular polarization in CMB radiation, so that the power spectrum of circular polarization of CMB $C_l^{V(S)}$ is proportional to the power spectrum of temperature anisotropy of CMB $C_l^{I(S)}$ and also $\delta^2$ which is a fraction of polarized electron number
density to the total one with net Left- or Right-handed polarizations. We have discussed that at least we need $\delta<10^{-4}$ to find consistency with a reported upper limit of CMB circular polarization.
\end{abstract}
\newpage
\section{Introduction}
CMB's temperature anisotropies which were detected by COBE in 1992 are believed to result from inhomogeneities in the matter distribution at the recombination epoch \cite{cobe}. Because Thomson scattering is an isotropic process, any primordial anisotropies (as opposed to inhomogeneities) should have been smoothed out before decoupling \cite{Gaw}. This certifies to the interpretation of the observed anisotropies as the result of density perturbations which can be a source for the formation of galaxies and clusters. The  temperature anisotropies which is discovered by COBE can be taken as evidence that such density inhomogeneities existed in the early universe \cite{Gaw,kosowsky19991,kosowsky19992}. Gravitational collapse of these primordial density inhomogeneities appears to have formed the large-scale structures of clusters,  super-clusters and galaxies which is observed today \cite{Gaw}.\\
Due to the anisotropic Compton
scattering around the recombination epoch, the generation of some relevant linear
polarization (about 10 percents) in CMB radiation is expected \cite{cosowsky1994,zal,hu}, and these polarization fluctuations
should be smaller than the temperature fluctuations \cite{nature}.
Most attractive results of Planck on cosmological parameters which included $r$ (scalar-tensor ratio) and linear polarization map is reported in \cite{planck, planck1,planck2}. On the other hand, according to the standard scenario of cosmology (considering Compton scattering as the mean interaction of CMB and cosmic matter), there is no physical mechanism to generate a circularly polarized radiation at the last scattering surface. It should be note that circular polarization measurements can provide valuable information to test the standard cosmological model
and the physics beyond the standard model of elementary particles. But experimental results confirm that one can still have circular polarization contribution in CMB anisotropy. Yet, there are relatively few published limits on the CMB circular polarization \cite{exp}. Almost all the experimental results have reported an upper limit for circular polarization (V-mode) around $\Delta_V/T_{CMB}<10^{-4}$. \\
Many reasons can be provided as to the generation of the circular polarization. In the case of  a renormalizable and gauge-invariant standard model, extension of photon coupling to an external vector field via Chern-Simons term, which arises as a radiative correction, if gravitational torsion couples to fermions, will be the source of circular polarization of CMB radiation \cite{Alexander}. The linear polarization of the CMB in the presence of a large-scale magnetic field B can be converted to the circular polarization under the formalism of the generalized Faraday rotation (FR)\cite{Jones, Cooray} known as the Faraday conversion (FC). Also, the V-mode can be produced with the same mechanism \cite{Massim, Massimo}. In a background magnetic field or the quantum electrodynamics sector of standard model which is extended by Lorentz non-invariant operators as well as non-commutativity, the CMB polarization acquires a small amount of circular polarization \cite{Bavarsad}. Photon-photon interactions mediated by the neutral hydrogen background, $\gamma+\gamma+atom \rightarrow \gamma+\gamma+atom$, through forward scattering \cite{Sawyer} and photon-neutrino scattering \cite{Mohammadi} and Euler-Heisenberg effective Lagrangian given in\cite{Euler} can produce the circular polarization. Also see other interesting mechanisms (like photon-graviton interaction, Magneto-optic effects,...) \cite{other}. \\
The scattering of a photon from the polarized electron is another mechanism which can be important for generating circular polarization. The description of polarization phenomena for both electromagnetic radiation and elementary particles as a matrix representation of polarization is given \cite{McMaster}. Some works are investigated phenomena involving electrons and photons from polarization effects considerations such as Compton scattering, bremsstrahlung and so on \cite{Lipp, Lipp2}. They studied the detection and production of circular polarized gamma radiation by Compton scattering. The production of polarized electron by photoionization of the polarized atomic beam has been reported which is the useful source of polarized electrons\cite{Long}. A Compton scattering based polarimeter for measuring the linear polarization of hard X-rays (100-300 keV) from astrophysical sources has been developing \cite{Mc}. A Monte Carlo method is described for the multi-pole scattering of linearly polarized gamma rays in non-magnetized solid state targets and then the cross section and Stokes parameters for spin-polarized have been discussed \cite{Bell}. Also, the investigation of $ \gamma- $ray polarizations leads to the insertion of the constraints on Planck scale violation of special relativity\cite{Kirk}. The final electron polarization was calculated for the scattering of the polarized photon by a polarized electron\cite{Kotkin}.\\
In this work, it is shown that Compton scattering of photons from polarized electrons\footnote{We call this in the remain of paper as “Polarized Compton Scattering”. } can generate circular polarization in contrast to the ordinary Compton scattering \cite{cosowsky1994}. The asymmetry between left- and right-handed number density of electrons can be obtained from several sources. For example, beta decays in Neutron stars \cite{neutronstar} as it is clearly shown, nature just accepts left-handed neutrinos and left-handed electrons flux intends in beta decay but their right-handed partner remains. Another one, axions as one of the dark-matter candidates can couple to fermions during inflation and produce both two helicity states of the electron but in asymmetrical amounts \cite{axion}. In the presence of magnetic field, electron should fill Landau levels \cite{landau}, while lowest Landau level can be filled only by left-handed electrons, higher levels filled with both helicity states of electrons. This will cause an asymmetry between left and right-handed electrons distribution which is in the order of $ \sim \dfrac{e B}{p^2} $, where $ B $ is the amplitude of magnetic field and $p$ is the linear momentum of electrons. By reviewing the chiral magnetic instability for electrons with only electromagnetic interaction, the chiral charge density, $ n_{5} $ is of order $ \backsimeq 10^{-14} n_{e} $ and $ n_{e} $ is the number density of electrons\cite{landau2}.
Also, electromagnetic interaction of the massive spin-$ 1/2 $ Dirac particles can flip their helicity \cite{Accioly}.
The above mentioned mechanisms motivated us to investigate the circular polarization generation of CMB via polarized Compton scattering.\\
\section{CMB Interaction with Polarized Electrons }
To describe an assumable of a photon like CMB radiation, one can start with the density matrix:
\begin{eqnarray}\label{rho}
\hat{\rho}=\dfrac{1}{tr(\hat{\rho})}\int \dfrac{d^{3}k}{(2\pi)^{3}}\rho_{\rm ij}(\bold{k})D_{\rm ij}(\bold{k})\end{eqnarray}
where $ D_{\rm ij} (\bold{k})\equiv a_{\rm i}^{\dagger}(\bold{k}) a_{\rm j}(\bold{k}) $ and $ \rho_{\rm ij} $ are the photon number operator and the general density-matrix component in the space of polarization states and $ \bold{k} $ indicates the momentum of photons. $I$, $Q$, $U$ and $V$ are Stokes parameters which is related to $\rho_{\rm ij}(\bold{k})$ as following
\begin{equation}
\hat{\rho}=\frac{1}{2}\left(
\begin{matrix}
I+Q &\,\, U-iV \\
U+iV&\,\, I-Q \\
\end{matrix}
\right)\label{matrix}
\end{equation}
The time evolution of $\rho_{\rm ij}(\bold{k})$ as well as Stokes parameters is given \cite{cosowsky1994},
\begin{eqnarray}\label{h0}
(2\pi)^3 \delta^3(0)2k^0
\frac{d}{dt}\rho_{\rm ij}(\bold{k}) \!\!&=&
i\langle[H^0_{\rm I}(t),D^0_{\rm ij}(\bold{k})]\rangle-\frac{1}{2}\int dt\langle\left[H^0_{\rm I}(t),[H^0_{\rm I}(0),D^0_{\rm ij}(\bold{k})]\right]\rangle
\end{eqnarray}
where $k^0=|\bold{k}|$ and $H^0_{\rm I}(t)$ is the first order of the interacting Hamiltonian.
The first term on the right-handed side of eq.(\ref{h0}) is a forward scattering term, and the second one is a higher order collision term.
Using standard calculations of Quantum Electrodynamics (QED), interacting Hamiltonian for electron-photon scattering ($\gamma(p)+e(q)\rightarrow \gamma(p')+e(q')$) is given
\begin{eqnarray}\label{h0interaction}
H^0_{\rm I}(t)=\int d\bold{q}d\bold{q'}d\bold{p}d\bold{p'}(2\pi)^3\delta^3(\bold{q'}+\bold{p'}-\bold{q}-\bold{p})exp{[it(q'^0+p'^0-q^0-p^0)]}\nonumber\\
\times[b^\dagger_{ r'}(\bold{q'})a^\dagger_{s'}(\bold{p'})\,\mathcal{M}(q'r',ps_1,qr,p's'_1)\,a_s(\bold{p})b_r(\bold{q})],
\end{eqnarray}
where $a_{\rm s}, a^{\dagger}_{\rm s'}$ and $b_{\rm r}, b^{\dagger}_{\rm r'}$ are annihilation and creation operators of the quantized photon and electron fields, respectively. $\mathcal {M}$ is Compton scattering amplitude
\begin{eqnarray}\label{12}
\mathcal{M}(q'r',ps_{1},qr,p's'_{ 1})=-ie^2\bar{U}_{\rm r'}(\bold{q'})\Big [\frac{\epsilon\!\!\!/_{\rm s'_1}({p'})\ (p\!\!\!/+q\!\!\!/+m)\epsilon\!\!\!/_{\rm s_1}(p)}{2q.p}-\frac{\epsilon\!\!\!/_{\rm s_1}(p)\ (q\!\!\!/-p'\!\!\!\!/+m)\epsilon\!\!\!/_{\rm s'_1}(p')}{2p'.q}\Big]U_r(\bold{q}),
\end{eqnarray}
where $r, r'$ and $s_1, s'_1$ indices run over electron and photon spin states. Note phase space elements are defined
\begin{eqnarray}\label{phase}
d\bold{q}=\frac{d^3\bold{q}}{(2\pi)^3}\frac{m}{q^0},~~~~~~~~~d\bold{p}=\frac{d^3p}{(2\pi)^32p^0}.
\end{eqnarray}

\subsection{Forward Scattering terms}
The usual assumption of forward scattering is that the fields begin as a free fields and end an other free field which the interactions are isolated from each other. According to this assumption, will prove (as it’s done by \cite{cosowsky1994})
\begin{eqnarray}
\bar{U}_r(q)\epsilon\!\!\!/_{s}(q\!\!\!/+m)\epsilon\!\!\!/_{s'}U_r(q)&&= \bar{U}_r(q)(2q\cdot\epsilon_s-q\!\!\!/\epsilon\!\!\!/_{s}+m\epsilon\!\!\!/_{s})\epsilon\!\!\!/_{s'}U_r(q)\nonumber\\
&&=(2q.\epsilon_s)\bar{U}_r(q)\epsilon\!\!\!/_{s'}U_r(q)\nonumber\\
&&=\frac{2}{m}(q\cdot\epsilon_s)(q\cdot\epsilon_{s'})\nonumber\\
&&=\bar{U}_r(q)\epsilon\!\!\!/_{s'}(q\!\!\!/+m)\epsilon\!\!\!/_{s}U_r(q)
\end{eqnarray}
These two terms (in Eq.(\ref{12})) cancel each other. Thus, the forward scattering of ordinary Compton scattering doesn't have any contribution to the generation of circular polarization of CMB's for electrons being unpolarized.
But this term in the presence of neutrino scattering off photons has a non-zero contribution \cite{rmohammadisadegh}.
The treating polarized electrons are the same as neutrinos or any other particles species governed by the Boltzmann equations.
With evaluation of the forward scattering term for Compton scattering, also this term will be zero independent of whether the electrons are polarized or unpolarized ( for more details see \cite{cosowsky1994} )
\subsection{Damping terms}
The contribution of damping term (usual cross section) of the Compton scattering for the generation of the CMB polarization has been studied in many works [see for example \cite{cosowsky1994} and its references]. At first, we just review the result presented in \cite{cosowsky1994} for the case of Compton scattering of photons from unpolarized electrons. The Boltzmann equation is
\begin{eqnarray}\label{eq:Boltz}
2k^0\dot{\rho}_{\rm ij}(\mathbf{k})&=&
\frac{1}{4}\int d\bold{q}d\bold{q'}d\bold{p}(2\pi)^4\delta^4(q'+p-q-k)
\mathcal{M}(q'r',ps'_1,qr,ks_1)\mathcal{M^{\dagger}}(qr,ks'_2,q'r',ps_2)\nonumber\\
&\times&{\Big[n_e(\bold x,\bold q)\delta_{\rm s_2\rm s'_1}(\delta_{\rm i\rm s_1}\rho_{\rm s'_2\rm j}(\mathbf{k})+\delta_{\rm js'_2}\rho_{\rm i\rm s_1}(\mathbf{k}))-2n_e(\bold x, \bold q')\delta_{\rm i\rm s_1}\delta_{\rm j\rm s'_2}\rho_{\rm s'_1\rm s_2}(\mathbf p)\Big]}
\end{eqnarray}
where $n_e(\bold x, \bold q)$ is the electron distribution function. The distribution function of cosmic electrons, which is known as a thermal Maxwell-Boltzmann distribution  \cite{cosowsky1994}, is
\begin{eqnarray}
n_e(\bold x,\bold q)=n_e(\bold x)(\frac{2\pi}{mT_e})^{3/2}exp{\big[-\frac{(\bold{\bold q}-m\bold{\bold v}(\bold x))^2}{2mT_e}\big]}
\end{eqnarray}
where $n_e(\bold x)$, $m$, $T_e$ and $\bold{\bold v}(\bold x)=v_e(\bold x)\hat{\bold v}$ are electron number density, electron mass, the electron temperature and the electron bulk velocity, respectively. Let's also write the following useful integrals
\begin{eqnarray}
&& \int \frac{d^3\bold q}{(2\pi)^3}n_e(\bold x,\bold q)=n_e(\bold x), \\
&& \int \frac{d^3\bold q}{(2\pi)^3}q_{\rm i} n_e(\bold x,\bold q)=m v_{\rm i}(\bold x)n_e(\bold x).
\end{eqnarray}

 Now with this furnishing, we go through the Eq.(\ref{eq:Boltz}). First we can simplify Compton scattering amplitude Eq.(\ref{12}) as follows
\begin{eqnarray}\label{simple12}
\mathcal{M}(q'r',ks_1,qr,ps'_1)=-ie^2\bar{U}_{\rm r'}(q')\Big [\frac{\epsilon\!\!\!/_{\rm s'_1}(p)(2q\cdot\epsilon_{\rm s_1}(k)-
	\epsilon\!\!\!/_{\rm s_1}(k)k\!\!\!/)}{2q\cdot k}-\frac{\epsilon\!\!\!/_{\rm s_1}(k)\ (2q\cdot\epsilon_{\rm s'_1}(p)+\epsilon\!\!\!/_{\rm s'_1}(p)p\!\!\!/)}{2p\cdot q}\Big]U_r(q)\nonumber\\
\end{eqnarray}
Then the squared Compton amplitude in abbreviated form is
\begin{eqnarray}
\mathcal{M}(q'r',ps'_1,qr,ks_1)\mathcal{M}(qr,ks'_2,q'r',ps_2)=e^4\sum\Big\{\bar U_{r'}(q')T(s_1,s'_1)U_r(q)\bar U_r(q)\tilde T(s_2,s'_2)U_{r'}(q')\Big\}, \label{SA}
\end{eqnarray}
where
\begin{eqnarray}
&&T(s_1,s'_1)=\frac{\epsilon\!\!\!/_{\rm s'_1}(p)}{2q\cdot k}[2q \cdot \epsilon_{\rm s_1}(k)-\epsilon\!\!\!/_{\rm s_1}(k)k\!\!\!/]-\frac{\epsilon\!\!\!/_{\rm s_1}(k)}{2q \cdot p}[2q \cdot \epsilon_{\rm s'_1}(p)+\epsilon\!\!\!/_{\rm s'_1}(p)p\!\!\!/]\label{ts1}\\
&&\tilde T(s_2,s'_2)=\frac{1}{2q\cdot k}[2q\cdot \epsilon_{\rm s'_2}(k)-k\!\!\!/\epsilon\!\!\!/_{\rm s'_2}(k)]\epsilon\!\!\!/_{\rm s_2}(p)-\frac{1}{2q\cdot p}[2q\cdot \epsilon_{\rm s_2}(p)+p\!\!\!/\epsilon\!\!\!/_{\rm s_2}(p)]\epsilon\!\!\!/_{\rm s'_2}(k)\label{ts2}
\end{eqnarray}
Note in the Compton scattering of unpolarized electrons, there are averaging assumption on the final and initial helicity states of electrons in Eq.(\ref{SA}), which allows to use the ordinary completeness relation $\sum_{r}U_{r}(\bold{q})\bar U_{r}(\bold{q})=\frac{q\!\!\!/’+m}{m}$ for both ingoing and out-coming electrons. But here we consider small polarization for ingoing electrons, in this case, the completeness relation of Dirac spinors modifies as \cite{kleiss}
\begin{eqnarray}\label{completeness}
U_{r}({q})\bar U_{r}({q})=\Big[\frac{q\!\!\!/+m}{2m}\frac{1+\gamma_5{S}\!\!\!/_r(\bold q)}{2}\Big]
\end{eqnarray}
where $S_r$ helicity operator with $r=L,R$ is defined
\begin{eqnarray}\label{HO}
S_R(\bold q)=(\frac{{\mid\bold q\mid}}{m},\frac{E}{m}\frac{\bold q}{\mid{\bold q}\mid}),~~~~~~~~S_L(\bold q)=-S_R(\bold q).
\end{eqnarray}
Let's consider a small fraction $\delta_L $ of left-handed polarization for ingoing cosmic electrons while we do not apply any constraint on the outgoing electrons due to their interaction with CMB photons. Hence, we have
\begin{eqnarray}\label{squared}
\mathcal{M}(q'r',ps'_1,qr,ks_1)\mathcal{M}(qr,ks'_2,q'r',ps_2)=e^4Tr\Bigg\{\frac{(q\!\!\!/'+m)}{2m_f}T(s_1,s'_1)\frac{(q\!\!\!/+m)}{2m_f}\Big[\frac{1+\gamma_5 {S}\!\!\!/_L(\bold q)}{2}\Big]\tilde T(s_2,s’_2)\Big]\Bigg\}
\end{eqnarray}
It should be noted that in the above equation $ q$ and $q'$ are ingoing and outgoing electrons momentum, respectively. One can rewrite Eq.(\ref{squared}) as following
\begin{eqnarray}\label{msquared}
\mathcal{M}(q'r',ps'_1,qr,ks_1)\mathcal{M}(qr,ks'_2,q'r',ps_2)&=&\frac{e^4}{2}Tr\Bigg\{\frac{q\!\!\!/'+m}{2m}T(s_1,s'_1)\frac{q\!\!\!/+m}{2m}\tilde{T}(s_2,s'_2)\Bigg\}\\
&+&\frac{e^4}{2}Tr\Bigg\{\frac{q\!\!\!/'+m}{2m}T(s_1,s'_1)\frac{q\!\!\!/+m}{2m}(\gamma_5{S}\!\!\!/_L(\bold q))\tilde{T}(s_2,s'_2)\Bigg\},\nonumber
\end{eqnarray}
where the first term is the amplitude of Compton scattering of unpolarized electrons which is investigated in standard scenario, whereas the second term indicates the contribution of Compton scattering of polarized electrons which is shown as $\mid\mathcal M\mid_{\rm P}^{~2}$.
With straightforward calculations (applicable Mathematica package\cite{mertig}) and keeping the dominated contribution, we have
\begin{eqnarray}
\mid\mathcal M\mid_{\rm P}^{~2}\approx \frac{e^4}{4(q\cdot k)^2}&\Bigg\{&q\cdot \epsilon_{\rm s'_2}(k)\Big(k\cdot \epsilon_{\rm s'_1}(p)\hat{q}\cdot \epsilon_{\rm s_1}(k)\times\epsilon_{\rm s_2}(p)+p\cdot \epsilon_{\rm s_1}(k)\hat{q}\cdot\epsilon_{\rm s'_1}(p)\times\epsilon_{\rm s_2}(p)\Big)\nonumber\\
&&+q\cdot\epsilon_{\rm s_2}(p)\Big(p\cdot\epsilon_{\rm s_1}(k)\hat{q}\cdot\epsilon_{\rm s'_2}(k)\times\epsilon_{\rm s'_1}(p)+\hat{q}\cdot\epsilon_{\rm s_1}(k)\epsilon_{\rm s'_2}(k)\cdot p\times\epsilon_{\rm s'_1}(p)\Big)\nonumber\\
&&+\hat{q}\cdot\epsilon_{\rm s'_1}(p)\Big(q\cdot\epsilon_{\rm s_2}(p)k\cdot\epsilon_{\rm s_1}(k)\times\epsilon_{\rm s'_2}(k)-q\cdot\epsilon_{\rm s'_2}(k)\epsilon_{\rm s_2}(p)\cdot k\times\epsilon_{\rm s_1}(k)\Big)\nonumber\\
&&-q\cdot\epsilon_{\rm s'_2}(k)\hat{q}\cdot\epsilon_{\rm s_1}(k)p\cdot\epsilon_{\rm s'_1}(p)\times\epsilon_{\rm s_2}(p)\nonumber\\
&&+\epsilon_{\rm s_1}(k)\cdot\epsilon_{\rm s'_1}(p)\Big(q\cdot\epsilon_{\rm s_2}(p)\hat{q}\cdot k\times\epsilon_{\rm s'_2}(k)-q\cdot\epsilon_{\rm s'_2}(k)\hat{q}\cdot k\times\epsilon_{\rm s_2}(p)\nonumber\\
&&+q\cdot\epsilon_{\rm s_2}(p)\hat{q}\cdot p\times\epsilon_{\rm s'_2}(k)-q\cdot\epsilon_{\rm s'_2}(k)\hat{q}\cdot p\times\epsilon_{\rm s_2}(p)\Big) \nonumber\\
&&+\epsilon_{\rm s_1}(k)\cdot\epsilon_{\rm s_2}(p)q\cdot\epsilon_{\rm s'_2}(k)\hat{q}\cdot p\times\epsilon_{\rm s'_1}(p)+\epsilon_{\rm s'_1}(p)\cdot\epsilon_{\rm s'_2}(k)q\cdot\epsilon_{\rm s_2}(p)\hat{q}\cdot k\times\epsilon_{\rm s_1}(k)\nonumber\\
&&-\delta_{\rm s_2\rm s'_1}q\cdot\epsilon_{\rm s'_2}(k)\hat{q}\cdot k\times\epsilon_{\rm s_1}(k)-\delta_{\rm s_1\rm s'_2}q\cdot\epsilon_{\rm s_2}(p)\hat{q}\cdot p\times\epsilon_{\rm s'_1}(p)\Bigg\},\label{mp1}
\end{eqnarray}
where $\hat{q}=\bold q/\mid q\mid$, then the Boltzmann equation for $\rho_{ij}(\bold{x},\bold{k})$ is given by
\begin{eqnarray}\label{cegamma}
\frac{d}{dt}\rho_{\rm ij}(\bold{x},\bold{k})&=&\frac{e^4\delta_L}{2k^0}\int d\bold{q}d\bold{p}\frac{m}{E(\bold{q}+\bold{k}-\bold{p})}(2\pi)\delta\big(E(\bold{q}+\bold{k}-\bold{p})+p-E(\bold{q})-k\big)\nonumber\\
&\times&\bigg(n_e(\bold x, \bold q)\delta_{\rm s_2\rm s'_1}(\delta_{\rm i\rm s_1}\rho_{\rm s'_2\rm j}(\mathbf{k})+\delta_{\rm j\rm s'_2}\rho_{\rm i\rm s_1}(\mathbf{k}))-2n_e(\bold x, \bold q')\delta_{\rm i\rm s_1}\delta_{\rm j\rm s'_2}\rho_{\rm s'_1\rm s_2}(\mathbf p)\bigg)\mid\mathcal M\mid_P^{~2}\nonumber\,,
\end{eqnarray}
where we introduce $\delta_{L}=n_{e,L}/n_e$ and $\delta_{R}=n_{e,R}/n_e$ as a fraction of polarized electron number density to total one with net Left- or Right-handed polarizations.By running all indices, ignoring the recoil momentum of final electrons and considering below equations
\begin{equation}\label{1}
\delta\big(E(\bold{q}+\bold{k}-\bold{p})+p-E(\bold{q})-k\big)\sim\delta\big(p-k\big),
\end{equation}
\begin{eqnarray}
E(\bold q+\bold Q)\sim m \big[1+\frac{\bold q^2}{m^2}+\frac{\bold q\cdot\bold Q}{m^2}+....\big]
\end{eqnarray}
\begin{eqnarray}
n_e(\bold q+\bold Q)\sim n_e(\bold q)\big[1-\frac{\bold Q\cdot(\bold q-m\bold v)}{m T_e}+....\big],
\end{eqnarray}
the time evolution of Stokes parameters would have the following form
\begin{eqnarray}\label{idotk}
\dot{I}(\mathbf{k})=\dot{\tau}_{_{\rm PC}}\int \frac{d\Omega}{4\pi}\sum_{_{\rm S}}\bigg[ f_{IS}{(\hat k,\hat p)}S(\bold k)+g_{IS}{(\hat k,\hat p)}S(\bold p)\bigg],
\end{eqnarray}
\begin{eqnarray}\label{qdotk}
\dot{Q}(\mathbf{k})=\dot{\tau}_{_{\rm PC}}\int \frac{d\Omega}{4\pi}\sum_{_{\rm S}}\bigg[ f_{QS}{(\hat k,\hat p)}S(\bold k)+g_{QS}{(\hat k,\hat p)}S(\bold p)\bigg]
\end{eqnarray}
\begin{eqnarray}\label{udotk}
\dot{U}(\mathbf{k})=\dot{\tau}_{_{\rm PC}}\int \frac{d\Omega}{4\pi}\sum_{_{\rm S}}\bigg[ f_{US}{(\hat k,\hat p)}S(\bold k)+g_{US}{(\hat k,\hat p)}S(\bold p)\bigg]
\end{eqnarray}
\begin{eqnarray}\label{vdotk}
\dot{V}(\mathbf{k})=\dot{\tau}_{_{\rm PC}}\int \frac{d\Omega}{4\pi}\sum_{_{\rm S}}\bigg[ f_{VS}{(\hat k,\hat p)}S(\bold k)+g_{VS}{(\hat k,\hat p)}S(\bold p)\bigg],
\end{eqnarray}
where $S\in\{I,Q,U,V\}$ and 
\begin{equation}\label{optical-pc1}
\dot{\tau}_{_{\rm PC}}=\frac{3}{2} \frac{m\,v_e(\bold x) }{k^0} \sigma_T\,\delta_L\,n_e(\bold x)\,.
\end{equation}
All coefficients $f_{IS}$, $f_{QS}$, $f_{US}$ and$f_{VS}$ can be easily obtained from Eqs.(\ref{matrix}) and (\ref{mp1}). As we are interested in calculation of the circular polarization (i.e. Eq.\eqref{vdotk}), we disregard the time evolution of $ \dot{I}(\mathbf{k}) $, $ \dot{Q}(\mathbf{k}) $ and $ \dot{U}(\mathbf{k}) $ for the rest. Also in (\ref{vdotk}), the coefficients of $ f_{VQ}$, $ g_{VQ}$, $ f_{VU}$ and $ g_{VU} $ are not considered because $ Q $ and $ U $ are at least one order of magnitude smaller than $ I $ in the case of CMB radiation.
\section{Power Spectrum of the Circular Polarization}
We continue the calculation in the presence of the primordial scalar perturbations indicated by $(S)$ which we expand in the Fourier modes characterized by a wave number $\mathbf{K}$. For each given
wave number $\mathbf{K}$, it is useful to select a coordinate system
with $\mathbf{K} \parallel \hat{\mathbf{z}}$ and
$(\hat{\mathbf{e}}_1,\hat{\mathbf{e}}_2)=(\hat{\mathbf{e}}_\theta,
\hat{\mathbf{e}}_\phi)$. The baryon bulk velocity $v$ at linear order is  irrotational,
meaning that it is the gradient of a potential, and thus in Fourier space it is parallel to wave number $\bold{K}$ (see \cite{hu2000}),
\begin{equation}\label{BV}
\bold{v}||\bold{K}\,\, , \,\,\,\,\,\,v=|\bold{v}|\approx(1+z)^{-1/2}10^{-3}.
\end{equation}
Temperature anisotropy $\Delta^{(S)}_{I}$ and circular polarization $\Delta^{(S)}_{V}$ of the CMB radiation can be expanded in the conformal time $\eta$ and can be described by multi-pole moments as following
\begin{eqnarray}
\Delta_{I,V}(\eta,\mathbf{K},\mu)=\sum^{\infty}_{l=0}(2 l+1)(-i)^l\Delta^l_{I,V}(\eta,\mathbf{K})P_{l}(\mu)
\end{eqnarray}
where $\mu = \hat{n}\cdot\hat{\mathbf{K}} = \cos \theta$, the $\theta$ is the angle between the CMB photon direction $\hat{n} = \mathbf{k}/|\mathbf{k}|$ and the wave vectors
$\mathbf{K}$ and $P_l(\mu)$ is the Legendre polynomial of rank $l$.
So, we could continue with the definition\footnote{This is confusing in the literature, but we should note that the right side of $\Delta^{(S)}_I$ is dimensionless and we continue with it.}
\begin{eqnarray}
\Delta^{(S)}_{I}(\mathbf{K},\mathbf{k},\eta)\equiv\left(4k\frac{\partial I_0}{\partial k}\right)^{-1} \Delta^{(S)}_ I(\mathbf{K},\mathbf{k},\eta).
\end{eqnarray}
Here we should define $\frac{d}{dt}$ in the left hand side of Eq.(\ref{cegamma}) to take into account space-time structure and gravitational effects such as the red-shift and so on. For each plane wave, each scattering and interaction can be described as the transport through a plane parallel medium \cite{mukh,chandra}, and finally, Boltzmann equations in the presence of the primordial scalar perturbations are given as
\begin{eqnarray}
&&\frac{d}{d\eta}\Delta _{V}^{(S)} +iK\mu \Delta _{V}^{(S)} = -\dot\tau_{e\gamma}\Big[\Delta _{V}^{(S)}-\frac{3}{2}\mu \Delta _{V1}^{(S)}\Big]-i2/3\dot{\tau}_{pc}\Big[P_2(\mu)\Delta_I^{(S)}-\Delta_{I2}^{(S)}\Big]
\label{Boltzmann2}
\end{eqnarray}
where $\dot{\tau}_{e\gamma}\equiv \frac{d\tau_{e\gamma}}{d\eta}$ which $\tau_{e\gamma}$ is Compton scattering optical depth, $a(\eta)$ is normalized scale factor.\\
The values of
$\Delta _{I}^{ (S)}(\eta_0,\hat{n})$ and $\Delta _{V}^{(S)}(\eta_0,\hat{n})$ at
the present time $\eta_0$ and the direction $\hat{n}$ can be obtained in the following general form by integrating of the Boltzmann equation (\ref{Boltzmann2}), along the line of sight \cite{zal} and with summing over all the Fourier modes $\mathbf{K}$,
\begin{eqnarray}
\Delta _{V}^{ (S)}(\hat{\bf{n}})
&=&\int d^3 \bf{K} \xi(\bf{K})\Delta _{V}^{(S)}
(\mathbf{K},\mathbf{k},\eta_0),\,\,\,\,\,\label{Boltzmann3}
\end{eqnarray}
where $\xi(\mathbf{K})$ is a random variable used to
characterize the initial amplitude of each primordial scalar perturbations mode, and then the values of
$\Delta _{V}^{(S)}(\mathbf{K},\mathbf{k},\eta_0)$ are given as
\begin{eqnarray}
\Delta _{V}^{(S)}
(\mathbf{K},\mu,\eta_0)
&\approx&\int_0^{\eta_0} d\eta\,
\dot\tau_{e\gamma}\,e^{ix \mu -\tau_{e\gamma}}\,\,\Big[ \frac{3}{2}\mu\Delta _{V1}^{(S)}-i\frac{2\dot\tau_{e\gamma}}{3\dot\tau_{pc}}(P_2(\mu)\Delta_I^{(S)}-\Delta_{I2}^{(S)})\Big],\label{VS}
\end{eqnarray}
where $x=K(\eta_0 - \eta)$. The differential optical depth $\dot\tau_{e\gamma}(\eta)$ and total optical depth $\tau_{e\gamma}(\eta)$ due to the Thomson scattering at time $\eta$ have been defined as follows
\begin{equation}\label{optical}
\dot{\tau}_{e\gamma}=a\,n_e\,\sigma_T,\,\,\,\,\,\,\,\tau_{e\gamma}(\eta)=\int_\eta^{\eta_0}\dot{\tau}_{e\gamma}(\eta) d\eta.
\end{equation}
The power spectrum  $ C_{l}^{V(S)} $, due to the Compton scattering in the presence of scalar perturbation, is
\begin{eqnarray}
C_{l}^{V(S)}=\langle \Delta _{Vl}^{(S)\dagger}
\Delta _{Vl}^{(S)}\rangle.
\end{eqnarray}
Therefore, the circular power spectrum of the CMB radiation, $ C_{l}^{V} $, due to Compton scattering would be
\begin{eqnarray}
C_{l}^{V}&=&\langle a_{Vl}^{\ast}a_{Vl}\rangle \nonumber\\
&\approx & \dfrac{1}{2l+1} \int d^{3} \bold{K} P_{\varphi}^{(S)}(\bold{K},\tau)\int \vert d\Omega P_{l}^{\ast} \int_{0}^{\tau_{0}}d\tau \dot{\tau}_{e\gamma} e^{ikx-\tau_{e\gamma}} [ \frac{2\dot{\tau_{pc}}}{3\dot\tau_{e\gamma}}(P_2(\mu)\Delta_I^{(S)}-\Delta_{I2}^{(S)}) ] \vert^{2}.\nonumber\\
\end{eqnarray}
and with the approximation, the power spectrum of circular polarization for $l<2$ can be estimated as
\begin{eqnarray}
C_{l}^{V (S)}\approx(\dfrac{\dot{\tau}_{PC}}{\dot{\tau}_{e\gamma}}\Big|_{\rm av})^{2}C_{l}^{I}=10^{8}\delta_L^2\,C_{l}^{I(S)},
\end{eqnarray}
where $C_{l}^{I(S)}$ is the power spectrum of temperature fluctuation and also, using Eq.(\ref{BV}), we have
\begin{equation}\label{av}
\dfrac{\dot{\tau}_{PC}}{\dot{\tau}_{e\gamma}}\Big|_{\rm av}\simeq\frac{m v_{e0}}{k^0}\frac{\delta_L}{z_{lss}}\int_0^{z_{lss}}\frac{dz}{(1+z)^{3/2}}\approx10^{4}\delta_L,
\end{equation}
where $ v_{e0}$ is the bulk velocity at present time and $z_{lss}$ indicates red-shift at last scattering surface.
\section{Conclusion}
In this work, according to assumption regarding an asymmetry in the number density of left- and right-handed electrons in the universe, we were motivated to calculate dominated contribution of this asymmetry for power spectrum of circular polarization $C_l^{V(S)}$ in CMB radiation. It should be mentioned that we have learned from papers \cite{McMaster}-\cite{Kotkin} to do our calculations. We have used Quantum Boltzmann Equation approach. The forward scattering term of polarized Compton scattering has no contribution to the CMB polarizations. We have shown that the damping term of polarized Compton scattering in the presence of scalar perturbation can generate circular polarization in CMB radiation, so that $C_l^{V(S)}$ is proportional to $C_l^{I(S)}$ and $\delta_L^2$. As our results showed, to generate circular polarization for CMB, the bulk velocity of cosmic electrons should be none zero. An interesting point is the converting anisotropy intensity $\Delta_I$ to circular polarization. The most observational groups have reported an upper limit around $\Delta_V/T_{CMB}<10^{-4}\sim \Delta T/T_{CMB}$ which means $C_l^{V(S)}\leq C_l^{I(S)}$. If we apply this upper limit, the fraction of polarized electron number density to the total one should be less than $\delta_L<10^{-4}$. As Eqs.(\ref{idotk})-(\ref{udotk}) shown, the polarized Compton scattering can generate the B-mode polarization in the presence of the scalar perturbation and can affect the value of E-mode polarization and the anisotropy of CMB temperature \cite{next}.


\end{document}